\documentclass[aps,prb,twocolumn]{revtex4} 

\usepackage{graphicx}
\usepackage{dcolumn}
\usepackage{bm}
\usepackage{amsmath}  

\newcommand{\comment}[1]{}

\begin{document}
\renewcommand{\theequation}{\arabic{section}.\arabic{equation}}

\title{Further On the Fountain Effect in Superfluid Helium}


\author{Phil Attard}
\affiliation{
{\tt phil.attard1@gmail.com}}
\noindent {\tt  Projects/QSM22/Paper4/fountain2.tex}


\begin{abstract}
In the previous paper (Attard 2022d)
on the fountain pressure in superfluid helium,
it was shown that the experimentally confirmed expression of H. London (1939)
was thermodynamically equivalent to equality of chemical potential.
However this theoretical equivalence
was not reflected in the experimental data.
The problem has now been traced to errors in the enthalpy and entropy
derived from the measured heat capacity by Donnelly and Barenghi (1998).
In this paper the corrected thermodynamic data yields
almost exact agreement between the two expressions
and the measured fountain pressure.
A physical explanation is given for energy minimization as 
the principle that drives the fountain effect and superfluid flow more generally.
\end{abstract}


\maketitle

%
\section{Introduction}
\setcounter{equation}{0} \setcounter{subsubsection}{0}
%

The fountain effect
refers to the spurting of superfluid helium
from a heated open chamber
connected by a capillary to a chamber of liquid helium
maintained below the condensation temperature (Allen 1938, Balibar 2017).
Closing the heated chamber and measuring the fountain pressure is,
in conjunction with an equation due to H. London (1939),
a common experimental technique for obtaining the entropy
(Donnelly and Barenghi 1998).

The accepted explanation for the fountain effect
as an osmotic pressure (Tisza 1938, Balibar 2017)
is quantitatively inaccurate
and makes little sense (Attard 2022d).

H. London's (1939) expression for the fountain pressure
---that the temperature derivative of the fountain pressure
equals the entropy per unit volume---
is known to be quantitatively accurate.
I have criticized his derivation on thermodynamic grounds (Attard 2022d),
which raises the question of the status of the result,
and, importantly, its meaning.
What does it tell us about superfluid flow,
and how does it relate to the broader principles of thermodynamics?
I showed that the expression
was thermodynamically equivalent to equality of chemical potential
in the two chambers at different temperatures  (Attard 2022d).
Although equality of chemical potential arises from energy minimization,
it is not at all clear why this should be relevant in thermodynamics,
or why it should be the principle that drives superfluid flow.

In contrast, for an equilibrium system
the maximization of the total entropy
would be the relevant principle,
but this would lead to equality of chemical potential divided by temperature,
which is not equivalent to the H. London (1939) expression.
The present system is a non-equilibrium steady state system
for which there is no principle of first entropy maximization
(Attard 2012).
And there is nothing specifically superfluid about entropy maximization.

This paper begins by recapitulating (Attard 2022d)
({\bf 1}) that equality of chemical potential arises from energy minimization,
({\bf 2}) that this is thermodynamically equivalent to the H. London (1939)
expression for the fountain pressure,
and ({\bf 3}) that equality of fugacity arises from entropy maximization.
The new material shows
({\bf 4}) that 1 and 3 are borderline distinguishable by measurement,
({\bf 5}) that there are systematic thermodynamic errors
in the enthalpy and entropy data reported by Donnelly and Barenghi (1998),
({\bf 6}) that the corrected thermodynamic data confirms the equivalence of 1 and 2
and their agreement with the measured fountain pressure
(Hammel  and Keller  1961),
and ({\bf 7}) that the adiabatic nature of collisionless superfluid flow
is the physical basis for the principle of energy minimization that underlies 1.

This paper is an excerpt from Ch.~9 on superfluidity
in the forthcoming second edition of Attard (2023).
It uses without justification some concepts from that chapter and
the preceding Ch.~8 on Bose-Einstein condensation
and the $\lambda$-transition.
It is largely independent of the theory of high temperature superconductivity,
Ch.~10.

%
\section{Thermodynamic analysis} \label{Sec:TD-ftn}
\setcounter{equation}{0} \setcounter{subsubsection}{0}
%

\subsection{Formal analysis}

Following Attard (2022d),
consider two closed chambers of helium,
$A$ and $B$,
each in contact with its own thermal reservoir of temperature
$T_A$ and $T_B$,
and having pressure $p_A$ and $p_B$.
The chambers are connected by a capillary through which fluid can flow.
Chamber $A$ in practice is at the lower temperature,
and consists of saturated liquid and vapor.
As H. London (1939) points out,
in the optimum steady state
the pressure of the second chamber is a function of its temperature
and the pressure and temperature of the first chamber,
$p_B = p(T_B;p_A,T_A)$.

The result given by H. London (1939)
says that the derivative of the pressure of the second chamber
with respect to its temperature for fixed first chamber
equals the entropy density,
\begin{equation} \label{Eq:HLondon-dp/dT}
\frac{\mathrm{d}p_B}{\mathrm{d}T_B} = \rho_B s_B .
\end{equation}
Here $\rho$ is the number density and $s$ is the entropy per particle.

H. London (1939) purported to derive this result
using a work-heat flow cycle,
which derivation is dubious (Attard 2022d appendix~A).
(See also section~\ref{Eq:anti-Nernst}.)
Possibly H. London (1939) simply guessed this result
and worked backwards:
the left hand side has units of Boltzmann's constant per unit volume,
and the only thermodynamic quantity with those units is the entropy density.


To derive this from a general axiom,
one should focus on variational principles for extensive quantities,
since these supply the foundations for thermodynamics
(Attard 2002, 2012, 2022d section~IIA).
Consider therefore the total entropy,
which is equivalent to the negative of the free energy divided by temperature
(Attard 2002).
Hence maximization of the former
gives the same result as minimization of the latter.
The first axiom that might determine the fountain pressure
is that the total entropy of the total system is a maximum.
This is of course just the Second Law of Thermodynamics,
albeit applied to a non-equilibrium steady state system.
Since the systems are closed, the total entropy is
(Attard 2023 section~2.2.2)
\begin{equation}
S_\mathrm{tot} =
\frac{- F(N_A,V_A,T_A)}{T_A} -  \frac{F(N_B,V_B,T_B)}{T_B} ,
\end{equation}
where $N$ is the number, $V$ is the volume,
and $F$ is the Helmholtz free energy.
With the total number of helium atoms fixed, $N=N_A+N_B$,
its derivative is 
\begin{equation}
\frac{\partial S_\mathrm{tot}}{\partial N_A}
 =
\frac{-\mu_A}{T_A} +  \frac{\mu_B}{T_B} ,
\end{equation}
where $\mu$ is the chemical potential.
The maximum total entropy occurs when this is zero,
which gives the condition for the optimum steady state  as
\begin{equation} \label{Eq:mu/T}
\frac{\mu_A}{T_A} =  \frac{\mu_B}{T_B} .
\end{equation}
Since the fugacity is $z = e^{\beta \mu}$,
where $\beta = 1/k_\mathrm{B}T$,
this is equivalent to $z_A  =  z_B $
and it can be called the constant fugacity condition.

Measurements of the fountain pressure
involve two closed chambers held at different temperatures
by a heater and a cooler.
Hence it is a non-equilibrium steady state system.
Usually maximization of the first entropy plays no direct role
in determining the optimum state of such systems
(Attard  2012, 2023 chapter~3).
Also, there is nothing specifically superfluid about this result.

The energy is also an extensive variable,
and the second possible axiom is that the total energy is a minimum.
The energy of each chamber is a function
of its entropy, volume, and number,
$E_\mathrm{tot} = E(S_A,V_A,N_A) +  E(S_B,V_B,N_B)$
(see sections~2.2.4 and 2.3.4 of Attard (2023)).
In this case the derivative at fixed total $N$ is
\begin{equation} \label{Eq:dEtot/dNA}
\frac{\partial E_\mathrm{tot}}{\partial N_A}
 =
\mu_A  - \mu_B .
\end{equation}
The energy is minimized (Attard 2023 section~9.4.5) when
\begin{equation} \label{Eq:muA=muB}
\mu_A =  \mu_B .
\end{equation}
For the physical interpretation of this result,
it is important to note that the derivative is at constant entropy.
There is currently no principle of energy minimization
in thermodynamics or statistical mechanics.
In mechanics,
the force points toward the potential energy minimum (Newton 1687),
but even in this case the total energy is constant on a trajectory.
Further, mechanical laws on their own
have no direct application to thermodynamic or statistical systems.

This same result can be obtained by minimizing the simple sum of
free energies, Helmholtz or Gibbs.
But it is difficult to justify simply adding them together
without dividing each by its respective chamber temperature.

The chemical potential is the Gibbs free energy per particle,
$\mu = G(N,p,T)/N$ (Attard 2002).
The derivative of equation~(\ref{Eq:muA=muB}) with respect to $T_B$
at constant pressure and temperature of the first chamber,
and number of the second, is  (Attard 2002, 2023 section~2.3.1)
\begin{eqnarray}
0
& = &
\frac{\mathrm{d}(G_B/N_B)}{\mathrm{d}T_B}
\nonumber \\ & = &
\frac{\partial g_B}{\partial T_B}
+
\frac{\partial g_B}{\partial p_B}
\frac{\mathrm{d}p_B}{\mathrm{d}T_B}
\nonumber \\ & = &
 -s_B + v_B \frac{\mathrm{d}p_B}{\mathrm{d}T_B} ,
\end{eqnarray}
where $g$, $s$, and $v = \rho^{-1}$
are the Gibbs free energy, entropy, and volume per particle,
respectively.
This is the same as H. London's expression,
equation~(\ref{Eq:HLondon-dp/dT}).
(Since $T_B = T_A$ implies $\mu_B=\mu_A$,
the constant of integration  must be zero.)
Therefore equality of chemical potential, equation~(\ref{Eq:muA=muB}),
is thermodynamically equivalent to
H. London's expression for the fountain pressure,
equation~(\ref{Eq:HLondon-dp/dT}).

\subsection{Relationship between the two principles}

Differentiating the condition of
constant fugacity, equation~(\ref{Eq:mu/T}),
gives
\begin{equation}
0 =
\frac{1}{T_B} \frac{\mathrm{d} \mu_B}{\mathrm{d} T_B}
- \frac{ \mu_B}{T_B^2} .
\end{equation}

On the saturation curve the chemical potential divided by temperature
must be small and negative because the liquid is in equilibrium with a gas.
Using measured values for the enthalpy and the entropy
for $^4$He 
(Donnelly and Barenghi 1998),
the value at $T=1$\,K is
$\beta \mu^\mathrm{sat} = -1.26\times 10^{-3}$,
and at $T=2.15$\,K it is
$\beta \mu^\mathrm{sat} = -1.05\times 10^{-1}$.
The fugacity
for bosons must be bounded above by unity, $z < 1$,
otherwise the denominator of the momentum state distribution
would pass through zero.
(For the case of ideal bosons,
F. London (1938) showed that $z^\mathrm{id}\to 1^-$
below the $\lambda$-transition (Attard 2022a, 2023 section~8.1).)
Since the compressibility is positive,
and since the fountain pressure is greater than the saturation pressure,
on a fountain path one must have
$  \mu^\mathrm{sat}_B  \le  \mu_B < 0$, or
\begin{equation}
-1 \ll \beta_B \mu_B < 0 .
\end{equation}
This result may be confirmed using
measured fountain pressures (Hammel and Keller 1961)
and saturation data (Donnelly and Barenghi 1998).

This reduces the derivative above  to
\begin{equation}
\frac{\mathrm{d} \mu_B}{\mathrm{d} T_B}
=
{\cal O}(10^{-3}) k_\mathrm{B} .
\end{equation}
This says that the condition of constant fugacity
is equivalent to the condition of constant chemical potential
to within about one part in one thousand.
Closer to the $\lambda$-transition the difference is about one part in ten.
One can see from this that
there is some question whether current measurements
have the accuracy to distinguish the two principles.


In the present case it may be said
that the condition of chemical potential equality
is a rigorous mathematical consequence of energy minimization
at constant entropy.
But not all mathematical results have physical relevance.
Given the fact that energy minimization plays no role
in the usual thermodynamic systems,
if it is indeed the underlying principle for superfluid flow
then a physical argument or explanation is required (see the conclusion).

Equally, it must be conceded that there is no reason to believe
that maximizing the entropy is a principle relevant
to the present non-equilibrium steady state system
(Attard 2012, 2023 chapter~3).

The tests against measured data (see section~\ref{Sec:FtnP-results}),
and the arguments that can be made  (see the conclusion),
both favor the principle of energy minimization.
Perhaps in time it will become axiomatic
that this is the principle that drives superfluid flow.

\subsection{Inconsistency in the derivation of the H. London expression}
\label{Eq:anti-Nernst}

In addition to the previous criticisms
of H. London's  (1939) derivation (Attard 2022d appendix~A),
here is a new argument that it is unsound.
The point is not whether his expression is correct,
but whether it has been mathematically proved  to be an exact result.

An essential step in H. London's  (1939) derivation
is where he invokes the Nernst heat theorem
to conclude that the enthalpy divided by temperature
and the entropy both go to zero at absolute zero.
These mean that the chemical potential
must go to zero at absolute zero.
H. London  (1939) also assumes that there exists a continuous fountain
path to absolute zero from an arbitrary thermodynamic point.

Since the H. London  (1939) expression implies
that the chemical potential is constant on the fountain path,
equation~(\ref{Eq:muA=muB}),
these assumptions imply that it is zero everywhere on the fountain path,
which is demonstrably false for saturated $^4$He.
It follows therefore that if the H. London (1939) expression is exact,
then either the Nernst heat theorem is invalid,
or else that there does not exist a fountain path to absolute zero,
or both.
In any case, the failure of either of these assumptions
renders the derivation of H. London (1939) invalid.

Without assessing the Nernst heat theorem in detail,
I note that if H. London (1939) is correct
in that it implies that the chemical potential divided by temperature
vanishes at absolute zero,
then it would imply that the fugacity equalled unity at absolute zero.
But the fugacity is bound to be strictly less than unity,
at least for a system with single particle energy states
(F. London 1938, Attard 2022a, 2023 section~8.1).
Contrariwise, the number of bosons would be infinite.
Since $^4$He is dominated by ideal statistics
deep below the $\lambda$-transition,
(see sections~9.1 and 9.2 of Attard (2023)),
a finite-sized system must arguably violate the Nernst heat theorem.

Similarly, on a fountain path to absolute zero,
situating the high temperature chamber on the saturation curve
would require a negative pressure in the other chamber at absolute zero.
(The saturated vapor pressure of the high temperature chamber is relatively low,
but it must still be much greater than the pressure
in the  chamber at absolute zero.)
Conversely,
if one insists upon a stable thermodynamic state at absolute zero,
then this places a lower bound on the pressure in the high temperature chamber
that would exceed the saturation pressure.
In other words, not all thermodynamic state points lie on a fountain path
to a stable point at absolute zero.

%
\section{Comparison with experiment} \label{Sec:FtnP-results}
\setcounter{equation}{0} \setcounter{subsubsection}{0}
%

\subsection{Three expressions and dubious data}

In practice in experimental application
the H. London (1939) expression for the derivative of the fountain pressure,
equation~(\ref{Eq:HLondon-dp/dT}),
is integrated along the saturation curve
(Hammel  and Keller 1961),
\begin{equation} \label{Eq:HL-int}
p_B - p_A =
\int_{T_A}^{T_B} \mathrm{d}T'\;
\rho^\mathrm{sat}(T') s^\mathrm{sat}(T')   .
\end{equation}

Strictly speaking,
the integral for the fountain pressure should be evaluated
on the fountain path rather than the saturation path.
But corrections for this effect involve the thermal expansivity,
$\alpha \sim {\cal O}(10^{-3})$,
and amount to only about $-0.5$\% at the highest fountain pressures (Attard 2022d).

The second equation for the fountain pressure comes from
the equality of chemical potential,
equation~(\ref{Eq:muA=muB}).
This can be used to obtain the fountain pressure by writing
$\mu_B = \mu_B^\mathrm{sat} +(p_B-p_B^\mathrm{sat}) v_B^\mathrm{sat}$,
which holds for an incompressible liquid.
Rearranging gives
\begin{eqnarray} \label{Eq:PA-mu}
p_B - p_A & \approx &
p_B^\mathrm{sat} - p_A
+ \rho_B^\mathrm{sat} (\mu_B - \mu_B^\mathrm{sat})
\nonumber \\ & = &
p_B^\mathrm{sat} - p_A
+ \rho_B^\mathrm{sat} (\mu_A - \mu_B^\mathrm{sat}) .
\end{eqnarray}
Invariably the experimental measurements are performed
at saturation of chamber $A$,
and so all quantities on the right hand side,
including $\mu^\mathrm{sat} = h^\mathrm{sat} - T s^\mathrm{sat}$,
can be obtained from standard tables
such as those given by Donnelly and  Barenghi (1998).
It is emphasized that any difference between the fountain pressure given by
equation~(\ref{Eq:HL-int}) and that given by equation~(\ref{Eq:PA-mu})
must be due to experimental error,
and the difference between them gives a guide to the reliability
of the measurements.

The third equation for the fountain pressure comes from
the equality of fugacity, equation~(\ref{Eq:mu/T}).
Again using the incompressible liquid expression for
the departure of the chemical potential from its saturation value,
this gives for the fountain pressure
\begin{equation} \label{Eq:PA-mu/T}
p_B -   p_A =
p_B^\mathrm{sat} -  p_A
+ \rho_B^\mathrm{sat}
\big( T_B \mu_A /T_A - \mu_B^\mathrm{sat} \big)    .
\end{equation}

\begin{figure}[t]
\centerline{ \resizebox{8cm}{!}{ \includegraphics*{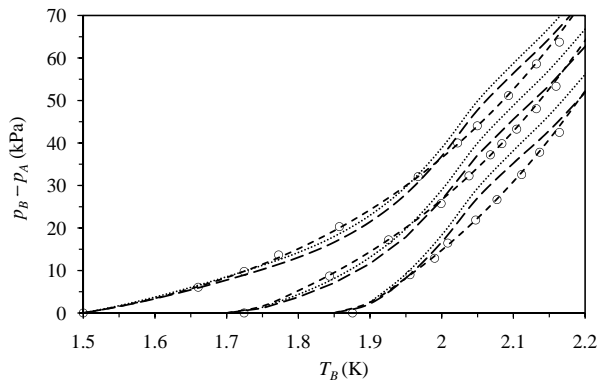} } }
\caption{\label{Fig:fountain1}
Measured and calculated fountain pressure for $T_A =$
1.502\,K (left), 1.724\,K  (middle), and 1.875\,K (right).
The symbols are measured data (Hammel and Keller 1961),
the short dashed curve is the saturation line integral form
of the H. London (1939) expression, equation~(\ref{Eq:HL-int}),
the dotted curve is for fixed chemical potential,
equation~(\ref{Eq:PA-mu}),
and the long dashed curve is for fixed fugacity,  equation~(\ref{Eq:PA-mu/T}).
The calculated curves use the  enthalpy (table 7.6)
and entropy (table 8.5)
obtained from measured heat capacity data
by Donnelly and Barenghi (1998).
}
\end{figure}

Figure~\ref{Fig:fountain1}
compares the three expressions for the fountain pressure
with the measured values of Hammel and Keller (1961).
This figure is essentially the same as that given in Attard (2022d).
Only the integral form
of the H. London (1939) expression, equation~(\ref{Eq:HL-int}),
agrees with the measured values.
What is somewhat concerning is that the results
for constant chemical potential, equation~(\ref{Eq:PA-mu}),
do not agree with the original H. London (1939) expression,
equation~(\ref{Eq:HL-int}), even though these are thermodynamically equivalent.
If the discrepancy were due to random experimental error
in either the thermodynamic data or the fountain pressure data,
then there should be a similar disagreement between the symbols
and the short dashed curve in the figure.

There must be an error,
either in my thermodynamic analysis,
or in the published experimental data used to check the expressions.
Until the error is identified and eradicated,
the data cannot be used to say which expression gives the fountain pressure,
and which principle, if either, drives superfluid flow.

\subsection{Corrected thermodynamic data}

The above test of these expressions against the measured fountain pressure
required measured values for the enthalpy per particle $h=H/N$,
the entropy per particle  $s=S/N$, and the chemical potential $\mu = h - Ts$.
Unfortunately the values given by Donnelly and Barenghi (1998)
are in error and need to be re-calculated.

The heat capacity at constant pressure is
(Attard 2012, 2023 section~2.4.2)
\begin{eqnarray}
C_\mathrm{p} & = &
\frac{-1}{T^2} \frac{\partial^2 (\overline G(N,p,T)/T)}{\partial (1/T)^2}
\nonumber \\ & = &
\frac{\partial \overline H(N,p,T)}{\partial T}
\nonumber \\ & = &
T \frac{\partial \overline S(N,p,T)}{\partial T} .
\end{eqnarray}
The second equality  can be seen by dividing both sides of the definition
of the unconstrained Gibbs free energy,
$G(E,V|N,p,T) = E + pV - TS(E,V,N)$, by $T$ and differentiating
with respect to $1/T$,
holding as usual $\overline E$ and $\overline V$ fixed.
The third equality follows by expressing the first equality
as a derivative with respect to $T$.

In practice measurements are made along the saturation curve,
$p^\mathrm{sat}(T)$.
I believe that the change in enthalpy,
$ \Delta H  = H(N,p^\mathrm{sat}(T_2),T_2) - H(N,p^\mathrm{sat}(T_1),T_1)$,
is measured as the energy input.
At constant number,
\begin{eqnarray} \label{Eq:dH/dT}
C_\mathrm{sat}
&\equiv &
\left( \frac{ \mathrm{d} \overline H(N,p,T) }{ \mathrm{d}T } \right)_N
\nonumber \\ & = &
\frac{\partial \overline H(N,p,T)}{\partial T}
+
\frac{\partial \overline H(N,p,T)}{\partial p}
\frac{\mathrm{d} p^\mathrm{sat}(T)}{\mathrm{d} T}
\nonumber \\ & = &
C_\mathrm{p}
+ \overline V \frac{\mathrm{d} p^\mathrm{sat}(T)}{\mathrm{d} T} .
\end{eqnarray}
This heat capacity at constant saturation
is larger than that at constant pressure.
I believe this to be the quantity
 $C_\mathrm{s}$ in table~7.4 of Donnelly and Barenghi (1998).

From this one sees that the difference in enthalpy on the saturation curve is
\begin{equation} \label{Eq:DH-sat}
H(N,p^\mathrm{sat},T)  -  H(N,p^\mathrm{sat}_0,T_0)
=
\int_{T_0}^T \mathrm{d} T' \; C_\mathrm{sat}(T').
\end{equation}
This contradicts the expression in note 11 to section~7
of Donnelly and Barenghi (1998),
who appear to have mixed up $C_\mathrm{p}$ and $C_\mathrm{sat}$.
The results they present for the enthalpy in their table~7.6
have a relative systematic error of ${\cal O}(10^{-2})$.
Although this is comparable to the random measurement and fitting error,
because it is a systematic error, and because the chemical potential
is the difference between two comparable quantities,
it leads to errors on the order of 5\% in the fountain pressure
(figure~\ref{Fig:fountain1}).

The temperature derivative of the entropy
along the saturation curve at constant number is
\begin{eqnarray} \label{Eq:dS/dT}
\lefteqn{
\left( \frac{ \mathrm{d} \overline S(N,p,T) }{ \mathrm{d}T } \right)_N
} \nonumber \\
& = &
\frac{\partial \overline S(N,p,T)}{\partial T}
+
\frac{\partial \overline S(\overline E, \overline V ,N)}{\partial p}
\frac{\mathrm{d} p^\mathrm{sat}(T)}{\mathrm{d} T}
\nonumber \\ & = &
\frac{1}{T} C_\mathrm{p}
-
\frac{\partial \overline V(N,p,T)}{\partial T}
\frac{\mathrm{d} p^\mathrm{sat}(T)}{\mathrm{d} T}
\nonumber \\ & = &
\frac{1}{T} C_\mathrm{p}
- \alpha  \overline V(N,p,T)
\frac{\mathrm{d} p^\mathrm{sat}(T)}{\mathrm{d} T}
\nonumber \\ & = &
\frac{1}{T} C_\mathrm{sat}
-
N\rho^{-1} \left[ \frac{1}{T} + \alpha  \right]
\frac{\mathrm{d} p^\mathrm{sat}(T)}{\mathrm{d} T}.
\end{eqnarray}
Accordingly the difference in entropy 
is
\begin{eqnarray} \label{Eq:DS-sat}
\lefteqn{
 S(N,p^\mathrm{sat}(T),T)  -  S(N,p^\mathrm{sat}(T_0),T_0)
}  \\ \nonumber
& = &
\int_{T_0}^T \mathrm{d} T' \; \frac{1}{T'}
\left\{
C_\mathrm{sat}(T')
-
\frac{N}{\rho'} \left[ 1 + \alpha' T' \right]
\frac{\mathrm{d} p^\mathrm{sat}(T')}{\mathrm{d} T'}
\right\} .
\end{eqnarray}
This contradicts the expression in note 8 to section~11
of Donnelly and Barenghi (1998),
which neglects the second term in the braces.
Compared to the present expression,
the results for the calorimetric entropy in their table~8.5
have a relative systematic error of ${\cal O}(10^{-2})$,
with a similar error for the fountain pressure
when used in the integral form
of the H. London (1939) expression, equation~(\ref{Eq:HL-int})
(figure~\ref{Fig:fountain1}).

\begin{figure}[t]
\centerline{ \resizebox{8cm}{!}{ \includegraphics*{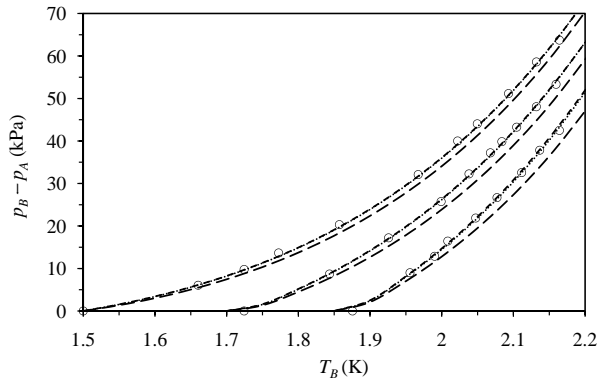} } }
\caption{\label{Fig:fountain2}
Same as the preceding figure, except that
the calculated curves use the measured saturation heat capacity,
$C_\mathrm{sat}$ (Donnelly and Barenghi 1998 table~7.4),
to obtain the enthalpy, equation~(\ref{Eq:DH-sat}),
and the entropy, equation~(\ref{Eq:DS-sat}).
}
\end{figure}

Figure~\ref{Fig:fountain2}
tests the various equations for the fountain pressure
against the measured values (Hammel and Keller 1961).
The calculations use the measured data for
the heat capacity at constant saturation,
$C_\mathrm{sat}$ (Donnelly and Barenghi 1998 table~7.4),
from which the enthalpy, equation~(\ref{Eq:DH-sat}),
and the entropy, equation~(\ref{Eq:DS-sat}),
and the chemical potential, $\mu = h - Ts$, are obtained.
This entropy was also used for the integrated H. London (1939) expression,
equation~(\ref{Eq:HL-int}).

It can be seen the fountain pressure predicted by
the H. London expression evaluated as an integral along the fountain curve,
equation~(\ref{Eq:HL-int}),
and that evaluated by equal chemical potential,
equation~(\ref{Eq:PA-mu}),
are virtually indistinguishable.
This confirms the thermodynamic equivalence of the two,
the validity of the thermodynamic analysis
that corrects the results of Donnelly and Barenghi (1998),
and the reliability of the experimental data when analyzed correctly.

It can be seen that the two forms of the  H. London expression
are in quite good agreement with the measured values
of the fountain pressure  (Hammel and Keller 1961).
Adding the correction for the translation
from the fountain path to the saturation path 
makes a difference of about $-0.5$\% at the highest fountain pressure shown,
which would be indistinguishable from the uncorrected result,
equation~(\ref{Eq:HL-int}), on the scale of the figure.

Hammel and Keller (1961) estimated the  error in their fountain pressure
measurements as about 2\%
and found that the predicted  fountain pressure, equation~(\ref{Eq:HL-int}),
using calorimetric entropy  values
(Kramers \emph{et al.}\ 1951, Hill and  Lounasmaa 1957)
agreed within the error.

For the heat capacity $C_\mathrm{s}=C_\mathrm{sat}$,
Donnelly and Barenghi (1998 table~7.4)
give the measurement error as 1--3\%
and the fitting error as 1--2\%.

The values of the fountain pressure predicted by equality of fugacity,
equation~(\ref{Eq:PA-mu/T}),
lie systematically below the measured values.
The difference  is on the order of 3--5\%,
which appears significant compared to the various measurement errors.
On the basis of the results in figure~\ref{Fig:fountain2},
one can tentatively conclude that the measured data
favor the principle of energy minimization at constant entropy
for superfluid flow,
and that they likely rule out the principle of entropy maximization.

\subsection{Convective flow}

In the fountain effect with closed chambers
a non-equilibrium steady state exists
with viscous flow of He~I from the high to the low temperature chamber
and
superfluid flow of He~II in the other direction to maintain mass balance.
These flows occur simultaneously in the same capillary.
There is also net energy flow
from the high temperature chamber to the low.
The flow of He~I is viscous Poiseuille flow
driven by the pressure difference (but see next)
and it carries the energy convectively
(F. London and Zilsel 1948, Keller and Hammel 1960).
The superfluid flow of He~II arriving in the high temperature chamber
is in total equal and opposite to the viscous flow of He~I leaving it.
A similar balance but in the opposite direction occurs
for the low temperature chamber.
In the steady state the total number of $^4$He atoms
in each of the two closed chambers is conserved.
There is of course a net energy flux between the two chambers,
with energy supplied by a heater in one
and removed by a refrigerator in the other.

In normal convective flow the two species are hot and cold particles
and their spatially distinct flows are driven by respective entropy gradients.
One would expect similar gradients in the present fountain system.
The evidence is that $\mu_A = \mu_B < 0$ (figure~\ref{Fig:fountain2}).
Since $T_B > T_A$, this means that $(-\mu_A/T_A) >  (-\mu_B/T_B)$,
which means that there is an entropy gradient that drives number
from $B$ to  $A$. (Recall $\partial S(E,V,N)/\partial N = -\mu/T$.)
This is what really drives the viscous Poiseuille flux of normal He, $J_*$,
from $B$ to $A$.

But what drives the steady  flow of condensed  He from $A$ to $B$?
Recall that when $\mu_A=\mu_B$ the energy is minimized
and there is no driving force.
One concludes that there must strictly be a gradient with
$\mu_B < \mu_A$, and, to linear order,  $J_0 =   c_2 (\mu_A-\mu_B)$.
In this case, then strictly speaking the measured fountain pressure
should lie between that predicted by constant chemical potential
and that predicted by constant fugacity.
The fact that the measured fountain pressure lies so close
to the prediction from $\mu_A=\mu_B$
indicates that superfluid flow is extremely efficient at eliminating gradients,
$c_2 \gg \beta_A J_*$.
It seems likely that the thinner the capillary,
and the lower the temperature difference,
the closer to equality would be the chemical potentials
(because the Poiseuille flow is reduced,
and a smaller balancing superfluid flow requires a smaller energy gradient).
It would appear that one needs a wide slit and better than the
2\% precision of current measurements
to confirm or refute the hypothesis $\mu_B < \mu_A$
(equivalently, $p_B^\mathrm{meas} < p_B^\mathrm{HLondon}$).


In the experiments of Keller and Hammel (1960),
the mean velocity of the viscous flow of He~I
in the case of the greatest fountain pressure
is on the order of 60 times the critical velocity for superfluid flow
predicted by the momentum gap for the slit
by the formula given in Attard (2023 section~9.3.4).
Assuming a comparable speed for the condensed bosons,
this suggests that collisions are strong enough to convert
a proportion of the back flow of superfluid He~II to viscous He~I.
But in the case of the fountain effect,
this does not block the capillary
because the fountain pressure is so large
that substantial Poiseuille flow continues.


%
\section{Conclusion} 
\setcounter{equation}{0} \setcounter{subsubsection}{0}
%

Experimental data is no substitute for a mathematical derivation
when it comes to proving that a result is exact.
But experimental data has historically been used
to formulate general scientific principles
that can then be used axiomatically to derive exact and approximate expressions
to describe that and other data.
It is usually the case that such principles gain acceptance over time
when no contradictory evidence emerges,
when they explain a range of physical phenomena,
and when scientists become familiar with them.
A rationale for the principle of energy minimization at constant entropy
for superfluid flow is now offered.

It was shown in section~\ref{Sec:TD-ftn} that H. London's (1939) expression
for the temperature derivative of the fountain pressure
corresponds to chemical potential equality of the two chambers.
It was also shown that
equal chemical potential minimizes the energy at constant entropy.

Equality of chemical potential
offers a thermodynamic mechanism for how the fountain pressure arises.
Imagine that initially the high temperature chamber is at saturation,
$\mu^\mathrm{ini}_B = \mu^\mathrm{sat}(T_B)$.
The low temperature chamber is always at saturation,
$\mu_A = \mu^\mathrm{sat}(T_A)$.
For $^4$He,
on the saturation curve the  chemical potential
from the measured enthalpy and entropy (Donnelly and Barenghi 1998)
decreases with increasing temperature,
$\mathrm{d} \mu^\mathrm{sat} /\mathrm{d} T < 0$.
Hence $\mu_B^\mathrm{sat} < \mu_A^\mathrm{sat}$.
Since $\partial \mu  /\partial p = v > 0$,
the only way that the chemical potential of the high temperature chamber
can be increased to achieve equality
is by increasing the pressure beyond its saturation value.
This occurs as more $^4$He arrives in the second chamber
because each atom occupies a certain impenetrable volume
(ie.\ the compressibility is positive).
Given the goal of equalizing the chemical potentials,
these are the reasons why condensed  $^4$He initially flows
down the chemical potential gradient from the low temperature chamber
to the high temperature chamber,
and why the high temperature chamber subsequently settles at a higher pressure.

The H. London (1939) expression,
and the thermodynamically equivalent equality of chemical potential,
agree with the measured values for the fountain pressure
within experimental error, figure~\ref{Fig:fountain2}.
This strongly suggests that energy minimization
is the principle that determines superfluid flow.
This raises ${\cal O}(10^0)$ questions:
Why is the principle of entropy maximization inapplicable?
Why is energy  minimization the operative principle?
And why is it at constant entropy?

In sections~9.1 and 9.2 of Attard (2023),
it is shown that the far side of the $\lambda$-transition
is dominated by the non-local permutation entropy
of bosons in multiply-occupied momentum states.
This entropy provides a barrier that suppresses momentum-changing collisions
(Attard 2022a).
Conversely, flow without such collisions conserves the permutation entropy.
These results show that superfluid flow is flow at constant permutation entropy.
And in so far as permutation entropy dominates the entropy of condensed $^4$He,
we may say that it is flow at constant entropy.
That superfluid flow does not change entropy explains why
the experimental data rules out conventional entropy maximization
as the underlying superfluid principle, figure~\ref{Fig:fountain2}.

Now consider a condensed boson
(ie.\ one in a multiply-occupied momentum state)
traversing the capillary from $A$ to $B$.
It must do so adiabatically (ie.\ with fixed total energy),
since it experiences no momentum-changing collisions.
Since it is condensed, this boson carries only mechanical energy $\mu_A$,
which is given by the derivative at constant entropy,
${\partial E(S,V,N)}/{\partial N}  = \mu$.
In contrast,
${\partial \overline E(N,V,T) }/{ \partial N } =
\mu - T { \partial \mu }/{ \partial T } $,
which has an entropic component.
On arriving in $B$,
our boson equilibrates via collisions to the temperature $T_B$,
the spacing of transverse momentum states being
much smaller in the chamber than in the capillary (Attard 2022d).
The change in entropy of chamber $B$ and its thermal reservoir
is the sum of that due to the change in energy, $\mu_A/T_B$,
and that due to the change in number, $-\mu_B/T_B$.
The change in entropy of chamber $A$ and its thermal reservoir
is the sum of that due to the change in energy, $-\mu_A/T_A$,
and that due to the change in number, $\mu_A/T_A$,
which cancel.
Hence the total change in total entropy
upon transfer of a condensed boson from $A$ to $B$
is $(\mu_A - \mu_B)/T_B$.
In the initial transient phase of the fountain effect,
$\mu_A > \mu_B$ (see above),
and so the change in the entropy of the universe is positive.
One concludes that the ultimate driving force for the fountain effect,
and superfluid flow more generally,
is entropy creation. Who knew?!

In summary,
superfluid flow transports energy adiabatically from
regions of high to low specific mechanical energy,
dissipating there the excess and creating entropy.
The flow is adiabatic
because superfluid flow is collisionless (Attard 2022a, 2023 chapter~9).
This explains the principle of energy minimization
as the immediate driving force for superfluid flow.

\section*{References}


\begin{list}{}{\itemindent=-0.5cm \parsep=.5mm \itemsep=.5mm}

\item 
Allen J F and Jones H 1938
New phenomena connected with heat flow in helium II
\emph{Nature} {\bf 141} 243

\item 
Attard  P 2002
\emph{Thermodynamics and statistical mechanics:
equilibrium by entropy maximisation}
(London: Academic Press)

\item 
Attard  P 2012
\emph{Non-equilibrium thermodynamics and statistical mechanics:
Foundations and applications}
(Oxford: Oxford University Press)


\item 
Attard P 2022a
Bose-Einstein condensation, the lambda transition, and superfluidity for
interacting bosons
arXiv:2201.07382

\item 
Attard P 2022d
On the fountain effect in superfluid helium
arXiv:2206.07914 [physics.chem-ph]

\item 
Attard  P 2023
\emph{Entropy beyond the second law.
Thermodynamics and statistical mechanics
for equilibrium, non-equilibrium, classical, and quantum systems}
(Bristol: IOP Publishing, 2nd edn)

\item 
Balibar S 2017
Laszlo Tisza and the two-fluid model of superfluidity
\emph{C.\ R.\ Physique} {\bf 18}  586

\item 
Donnelly R J and  Barenghi C F 1998
The observed properties of liquid Helium at the saturated vapor pressure
\emph{J.\ Phys.\ Chem.\ Ref.\ Data} {\bf 27} 1217

\item  
Hammel (Jr) E F  and Keller W E 1961
Fountain pressure measurements in liquid He~II
\emph{Phys.\ Rev.}\ {\bf 124} 1641

\item  
Hill R W and Lounasmaa O V 1957
\emph{Phil.\ Mag.}\  {\bf 2}, 1943

\item  
Keller W E  and Hammel (Jr) E F 1960
Heat conduction and fountain pressure in liquid He~II
\emph{Annals of Physics} {\bf 10}  202

\item 
Kramers H C, Wasscher J D and  Gorter C J 1951
\emph{Physica} {\bf  18} 329

\item  
London F 1938
The $\lambda$-phenomenon of liquid helium and the Bose-Einstein degeneracy
\emph{Nature} {\bf 141} 643

\item 
London H 1939
Thermodynamics of the thermomechanical effect of liquid He II
\emph{Proc.\ Roy.\ Soc.}\ {\bf  A171} 484

\item  
London F  and  Zilsel P R 1948
Heat transfer in liquid helium II by internal convection
\emph{Phys.\ Rev.}\ {\bf 74} 1148

\item 
Newton I 1687
\emph{Philosophi\ae\ Naturalis Principia Mathematica}
(London: S. Pepys)

\item  
Tisza L 1938
Transport phenomena in helium II
\emph{Nature} {\bf 141} 913

\end{list}




\end{document}